\documentclass[aps,prl,superscriptaddress,floatfix,showpacs,twocolumn]{revtex4}

\usepackage{verbatim}

\usepackage{amsmath}
\usepackage{amsfonts}
\usepackage{amssymb}
\usepackage{graphicx}
\usepackage{bm}
\usepackage{color}

\usepackage{epsf}

\usepackage{psfrag}

\def\bea{\begin{eqnarray}}
\def\eea{\end{eqnarray}}
\def\beas{\begin{eqnarray*}}
\def\eeas{\end{eqnarray*}}
\def\beqas{\begin{eqnarray*}}
\def\eqas{\end{eqnarray*}}
\def\beq{\begin{equation}}
\def\eeq{\end{equation}}
\def\beqd{\begin{displaymath}}
\def\eeqd{\end{displaymath}}
\def\eqd{\end{displaymath}}

\def\slashchar#1{\setbox0=\hbox{$#1$}
   \dimen0=\wd0
   \setbox1=\hbox{/} \dimen1=\wd1
   \ifdim\dimen0>\dimen1
      \rlap{\hbox to \dimen0{\hfil/\hfil}}
      #1
   \else\begin{eqnarray}
      \rlap{\hbox to \dimen1{\hfil$#1$\hfil}}
      /
   \fi}

\begin{document}
\title
{Soft-collinear resummation in deeply virtual Compton scattering }
\author{T. Altinoluk}
\author{ B.~Pire}
\affiliation{
CPhT, \'Ecole Polytechnique,
CNRS, F-91128 Palaiseau,     France}
\author{ L. Szymanowski}
\affiliation{National Center for Nuclear Research (NCBJ), Warsaw, Poland}
\author{S. Wallon}
\affiliation{LPT, Universit{\'e} Paris-Sud, CNRS, 91405, Orsay, France {\em \&} \\
UPMC Univ. Paris 06, facult\'e de physique, 4 place Jussieu, 75252 Paris Cedex 05, France}


\begin{abstract}

\noindent
We derive an all order resummation formula for the deeply virtual Compton scattering (DVCS) amplitude, which takes into account soft gluon exchanges in the non-singlet quark coefficient function. We identify the ladder diagrams responsible in light-like gauge for $(\alpha_s \log^{2}(x\pm \xi))^n$ contributions. The resummed series results in a simple closed expression. 
\end{abstract}

\maketitle

In the  collinear factorization framework the scattering amplitude for  exclusive processes such as DVCS
has been shown \cite{historyofDVCS} to factorize in specific kinematical regions, provided a large scale controls the separation of short distance dominated partonic subprocesses and long distance hadronic matrix elements, the generalized parton distributions (GPDs) \cite{review}. 
The  amplitude for the DVCS process
\begin{equation}
\gamma^{(*)}(q) N(p) \to \gamma(q') N'(p')\,.
\label{reaction}
\end{equation}
with a large  virtuality $q^2 =-Q^2$, factorizes in terms of perturbatively calculable coefficient functions $C(x,\xi, \alpha_s) $ and GPDs $F(x,\xi,t)$, where the scaling variable in the generalized Bjorken limit is the skewness $\xi$ defined as 
\begin{equation}
\xi = \frac{Q^2}{(p+p')\cdot(q+q')}\,.
\label{eq:skewnessdef}
\end{equation}

\vspace{.2cm}
\paragraph*{Lessons from the  DVCS amplitude at NLO.}
Let us begin with  the discussion of the NLO corrections to the  amplitude for  DVCS (\ref{reaction}). 
After proper renormalization, the quark contribution to the symmetric part
of the factorized Compton scattering amplitude reads 
\begin{eqnarray}
\mathcal{A}^{\mu\nu} = g_T^{\mu\nu}\int_{-1}^1 dx 
\left[
\sum_q^{n_F} T^q(x) F^q(x)
\right]\,,
\label{eq:factorizedamplitude}
\end{eqnarray}
where the  quark coefficient function $T^q$ reads \cite{Pire:2011st} 
\begin{eqnarray}
\!\!\!\!\!\!T^q&\!\!=&\!\! C_{0}^q +C_1^q +C_{coll}^q \log \frac{|Q^2|}{\mu^2_F}  \,,\\
\!\!\!\!\!\!C_0^q &\!\!=&\!\! e_q^2\left(\frac{1}{x-\xi+i\varepsilon} \,-\, (x \to -x)  \right) \,, \label{C0} \\
\!\!\!\!\!\!C_1^q &\!\!=&\!\! \frac{e_q^2\alpha_SC_F}{4\pi(x-\xi+i\varepsilon)}
\bigg\{
\log^2(\frac{\xi-x}{2\xi}-i\varepsilon )
\,-\,  9 \nonumber \\
&&- \, 3\frac{\xi-x}{\xi+x}\log\bigg(\frac{\xi-x}{2\xi}-i\epsilon\bigg)\bigg\}-(x \to -x) \,.
\label{C1}
\end{eqnarray}
The first (resp. second) terms in Eqs. ~(\ref{C0}) and (\ref{C1}) correspond to the $s-$channel (resp. $u-$channel) class of diagrams. One goes from the $s-$channel to the  $u-$channel by the interchange of the photon attachments. Since these two contributions are obtained from one another by a simple ($x\leftrightarrow -x$) interchange, we will restrict in the following mostly to the discussion of the former class of diagrams.

Eqs. ~(\ref{C0}) and (\ref{C1}) show that among the corrections of $O(\alpha_s)$ to the coefficient function,  the terms of order $\frac{\log^2(x\pm\xi)}{x\pm\xi}$ play an important role in the region of small $(x\pm\xi)$,  i.e. in the vicinity of the boundary between the so-called ERBL and DGLAP domains, where the evolution equations of GPDs take distinct forms. 

\begin{figure}
\includegraphics[width=8.5cm]{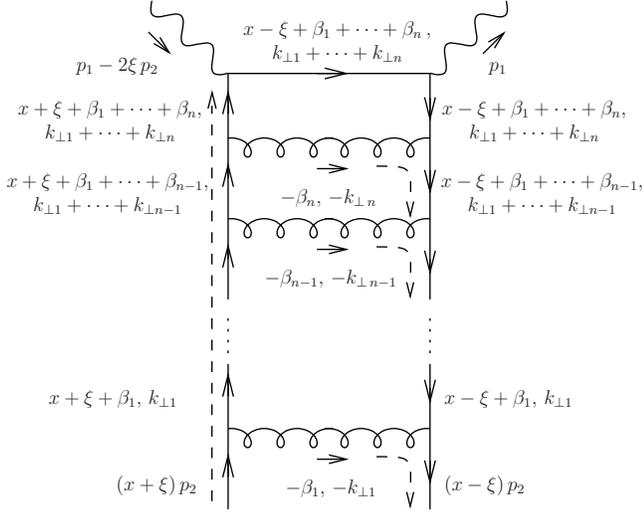}
\vspace{-2mm}
\caption{The ladder diagrams which contribute in the light-like gauge to the leading  $\alpha_s^n \ln^{2 n}(\xi-x)/(x-\xi)$ terms in the perturbative expansion of the    DVCS  amplitude. The $p_2$ and $\perp$ momentum components are indicated. The dashed lines show the dominant momentum flows along the $p_2$ direction.} 
\vspace{-4mm}
\label{Fig:n-loop}
\end{figure}

\vspace{.2cm}
\paragraph*{Main steps of our analysis.}
We start our analysis with the observation that in the same spirit as for evolution equations, the extraction of the soft-collinear singularities which dominate the amplitude in the limit $x \to \pm \xi$ is made easier if one uses the light-like gauge $p_1 \equiv p_\gamma$. We argue that in this gauge the amplitude is dominated by ladder-like diagrams, see Fig.~\ref{Fig:n-loop}. In our analysis we expand any momentum in the Sudakov basis $p_1$, $p_2$, 
as $k = \alpha \, p_1 + \beta \, p_2 + k_\perp\,,$
where $p_2$ is the light-cone direction of the two incoming and outgoing partons ($p_1^2=p_2^2=0$, $2 p_1 \cdot p_2 =s = Q^2/2\xi $).
 In this basis,  $q_{\gamma^*}=p_1-2 \, \xi \, p_2$\,. 

We now restrict our study to the limit   $x \to +\xi$. The dominant kinematics is given by a strong ordering both in longitudinal and transverse momenta, according to
\begin{eqnarray}
&& \hspace{-.3cm}x \sim \xi\gg \vert \beta_1\vert  \sim \vert x-\xi \vert \gg \vert x-\xi -\beta_1\vert \sim \vert \beta_2 \vert \gg \cdots \nonumber \\
&&\hspace{-.4cm}
\cdots \gg   \vert x-\xi -\beta_1 -\beta_2 -\cdots- \beta_{n-1} \vert  \sim  \vert \beta_n  \vert  , 
\label{kinematics_beta}
\\
&& \vert k_{\perp 1}^2\vert  \ll \vert k_{\perp 2}^2\vert  \ll \cdots \ll \vert k_{\perp n}^2\vert  \ll s \sim Q^2 \,,
\label{kinematics_k}
\\
&&  \vert \alpha_1 \vert  \ll \cdots \ll  \vert \alpha_n \vert  \ll 1\,.
\label{kinematics_alpha}
\end{eqnarray}
This ordering is related to the fact that the dominant double logarithmic contribution for each loop arises from the region of phase space where both soft and collinear singularities manifest themselves. When $x \to \xi$ the left fermionic line is a hard line, from which the gluons are emitted in an eikonal way, with a collinear ordering. For the right fermionic line, eikonal approximation is not valid, since the dominant momentum flow along $p_2$ is from  gluon to  fermion, nevertheless the collinear approximation can still be applied. 

Finally,
the issue related to the $i \epsilon$ prescription in Eq.~(\ref{C1}) is solved by computing the coefficient function in the unphysical region $\xi
> 1$. After analytical continuation to the physical region $0 \leq \xi \leq 1$, the final result is then obtained through the shift $\xi \to \xi-i \epsilon$\,.

To prepare for further analysis,
we define  $K_n$, the contribution of a $n$-loop ladder to the coefficient  function, as
\begin{eqnarray}
\label{def-Kn}
K_n = -\frac{i}4 e_q^2 \left(-i \, C_F \, \alpha_s \frac{1}{(2 \pi)^2} \right)^n I_n\,.
\end{eqnarray}
Now we will discuss in details the main points of derivation of $I_n$ starting from the simplest case.

\unitlength 1mm

\vspace{.2cm}
\paragraph*{The ladder diagram at order $\alpha_s$.}

\begin{figure}
\begin{tabular}{ccccc}
\ \includegraphics[width=1.4cm]{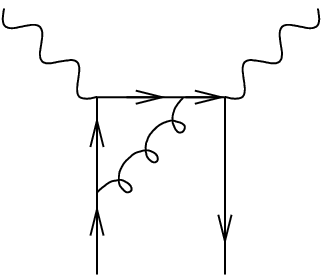} &
\qquad \raisebox{.01cm}{\includegraphics[width=1.55cm]{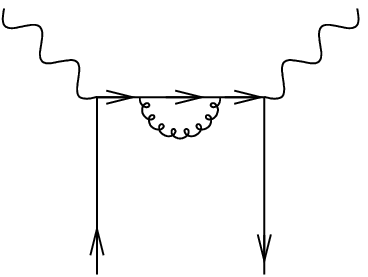}} &
\qquad \includegraphics[width=1.4cm]{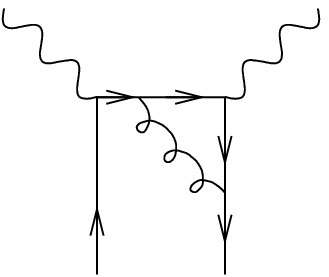} 
\end{tabular}
\caption{The one-loop diagrams which do not contribute to the leading  $\frac{\log^2(x-\xi) }{x-\xi}$ terms in the $p_1 \cdot A=0$ light-like gauge.}
\vspace{-4mm}
\label{Fig:suppressed-1loop}
\end{figure}

A careful analysis shows that the one-loop diagrams of Fig.~\ref{Fig:suppressed-1loop} are not dominant for $x \to \xi\,$. 
Thus, we concentrate on the box diagram, see Fig.~\ref{Fig:n-loop} with $n = 1$.
Starting from the dominant part of the numerator of the Born term which is $\not{\theta}=-2\not{p}_1$, 
the numerator of the box diagram is
\begin{equation}
tr\big\{ \not{p_2}\gamma^{\mu}[\not{k}+(x-\xi)\not{p_2}]\not{\theta}[\not{k}+(x+\xi)\not{p_2}]\gamma^{\nu} \big\}d^{\mu\nu}\,.
\end{equation}
We first perform the soft gluon approximation : $k+(x+\xi)p_2 \to (x+\xi)p_2$.
Since the gluon is almost on-shell, the dominant contribution in the gluon propagator, $d^{\mu\nu}$, when expressed in terms of gluon polarization vectors, is given by 
$
d^{\mu\nu}\approx -\sum_{\lambda}\epsilon^{\mu}_{(\lambda)}\epsilon^{\nu}_{(\lambda)}\,.
$
Writing gluon polarization vectors through their Sudakov decomposition
\begin{equation}
\epsilon^{\mu}_{(\lambda)}= \epsilon^{\mu}_{\perp (\lambda)} -2 \frac{\epsilon_{\perp (\lambda)}\cdot k_{\perp}}{\beta s}p_1^\mu \,,
\end{equation} 
allows us to define  an effective vertex for the gluon and outgoing quark through the polarization sum 
\begin{equation}
\sum_{\lambda}\epsilon_{\perp (\lambda)}\cdot k_{\perp}\epsilon^{\mu}_{(\lambda)}=\bigg(-k_{\perp}^ \mu+2\frac{k_{\perp}^2}{\beta s}p_1^\mu\bigg) \,.
\end{equation}
 The numerator, $(Num)_1$, is $\alpha-$independent and reads
\begin{eqnarray}
&&\hspace{-.7cm}\frac{-4(x+\xi)}{\beta}tr\bigg\{\!\!\not{p_2}\bigg(\not{k_{\perp}}-2\frac{k_{\perp}^2}{\beta s}\not{p_1}\bigg)[\not{k}+(x-\xi)\not{p_2}]\not{p_1}\!\!\bigg\} \nonumber \\
&=&-4(x+\xi)s\frac{2k_{\perp}^2}{\beta}\bigg[1+\frac{2(x-\xi)}{\beta}\bigg] \,.
\end{eqnarray}
We emphasize that the term $(x+\xi)/\beta$  arises from the eikonal emission from the left fermionic line while the expression inside the  $[ \,\cdots ]$ accounts for the fact that gluon is not soft from the point of view of the right fermion (in an eikonal treatment it would reduce to $(x-\xi)/\beta$).
%

Let us now calculate the  integral over the gluon momentum $k$,
using  dimensional regularization 
$  \int d^dk \rightarrow \frac{s}{2}\int d\alpha \,d\beta \,d^{d-2}\underline{k}$, $(k_{\perp}^2=-\underline{k}^2)$.
The Cauchy integration over $\alpha$, which also determines the limits of the $\beta$ integration, gives 
   \begin{eqnarray}
 & I_1& = - 2  \pi i \frac{s}{2} \bigg\{ \int_0^{\xi-x}\frac{d\beta}{s\beta }\int_0^{\infty}d^{d-2}\underline{k} \frac{(Num)_{1}}{L_1^2 \,R_1^2\, S^2}\Big{\vert}_{\alpha=\frac{\underline{k}^2}{\beta s}} \nonumber \\
  &+& \int_{-\xi-x}^{\xi-x}\frac{d\beta}{s\beta }\int_0^{\infty}d^{d-2}\underline{k} \frac{(Num)_{1}}{L_1^2 \,R_1^2\, S^2}\Big{\vert}_{\alpha=\frac{\underline{k}^2}{(\beta +x+\xi)s}} \bigg\}\,,
\label{oneloop}
 \end{eqnarray}
with the denominators of propagators  $L_1$($R_1$) on the left (right) quark line and $S$ for s-channel quark line
\begin{eqnarray}
L_1^2&=&\alpha(x+\xi)s ~~~,~~~R_1^2=-\underline{k}^2+\alpha(\beta +x-\xi)s \, , \nonumber \\
S^2&=&-\underline{k}^2+(\beta+x-\xi)s \,.
\end{eqnarray}
 It turns out that  the first line in Eq.~(\ref{oneloop}), which corresponds to cutting through the exchanged gluon, dominates. Moreover, the relevant region of integration corresponds to small $\vert \beta+x-\xi \vert $. The $\beta$ and $\underline{k}$ integrations  results in our final one-loop expression
\begin{equation}
\label{I1}
I_1=-\frac{4}{x-\xi}\frac{2  \pi i}{2!}\log^2(a(x-\xi)) \,,
\end{equation}
where we kept only the most singular terms in the $x\to \xi$ region and have no control of the value of $a$ within our approximation. To fix $a$, we match our approximated one-loop result with the 
full one-loop result (\ref{C1}). This amounts to cut the $\underline{k}^2$ integral at $Q^2$. The $i \epsilon$ term is included  according to the same matching. This leads to
\begin{equation}
\label{I1-final}
I_1=-\frac{4}{x-\xi+ i \epsilon}\frac{2  \pi i}{2!}\log^2\left(\frac{\xi- x}{2\xi}-i \epsilon\right) \,.
\end{equation} 
 A similar expression holds  for the $u-$channel diagram
in the limit $x\to -\xi\,.$

\vspace{.2cm}

\paragraph*{The ladder diagram at order $\alpha_s^2$.} Let us examine the next order in the perturbative expansion. There are many diagrams contributing  but it can be shown that in the chosen gauge, the double box diagram dominates, and moreover that the relevant region of integration  is where there is a strong ordering between $k_{2\perp}$ and $k_{1\perp}$, i.e. $k_{2\perp}\gg k_{1\perp}$. Using the same arguments as in the one-loop case, one  simplifies the numerator for this diagram, after  
taking advantage of strong $k_{i\perp}$ ordering, as
\begin{eqnarray}
(Num)_2&=&-4s(x+\xi)^2\bigg\{\frac{2k_{\perp1}^2}{\beta_1}\bigg[1+\frac{2(x-\xi)}{\beta_1}\bigg]\bigg\}\nonumber \\
&&\times\bigg\{\frac{2k_{\perp 2}^2}{\beta_2}\bigg[1+\frac{2(\beta_1+x-\xi)}{\beta_2}\bigg]\bigg\}\,.
\end{eqnarray}
The denominators of fermionic propagators read (Fig.~\ref{Fig:n-loop})
 \begin{eqnarray}
L_1^2&=& \alpha_1(x+\xi)s~~,~~R_1^2=-\underline{k}^2+\alpha_1(\beta_1+x-\xi)s \,,\nonumber \\
L_2^2&=&\alpha_2(x+\xi)s~~,~~R_2^2=-\underline{k}^2+\alpha_2(\beta_1+\beta_2+x-\xi)s \,,\nonumber \\
S^2&=&-\underline{k}^2+(\beta_1+\beta_2+x-\xi)s \,.
 \end{eqnarray}
 Using  Cauchy integration over $\alpha_1$ and $\alpha_2$, we see that  the dominant contribution comes from the residues corresponding to the two gluon propagators, and reads 
 \begin{equation}
\hspace{-.0cm}I_2\approx  \int_0^{\xi-x}\hspace{-.4cm}d\beta_1\int_0^{\xi-x-\beta_1}\hspace{-.6cm}d\beta_2\int_0^{\infty}\hspace{-.2cm}d^{d-2}\underline{k}_2\int_0^{\underline{k}_2^2}\hspace{-.2cm}d^{d-2}\underline{k}_1\,{\cal I}_2 \,,\end{equation}
 where  
 \begin{equation}
 \hspace{-.2cm}{\cal I}_2  \!\!=\!\!\frac{4\,s(2\pi i)^2}{x-\xi} \frac{1}{\beta_1+x-\xi}\frac{1}{\underline{k}_1^2}\frac{1}{\underline{k}_2^2}\frac{1}{\underline{k}_2^2-(\beta_1+\beta_2+x-\xi)s}.
\end{equation}
The integrals over $\underline{k}_1$ and $\underline{k}_2$ are performed by assuming that the singularities are regularized within dimensional regularization. Taking into account the fact that a scaleless integral vanishes, we get the following formula
\begin{equation}
\label{I2-final}
I_2\approx  -\frac{4}{x-\xi+i\epsilon} \frac{(2 i \pi)^2}{4!}\log^4\left(\frac{\xi-x}{2\xi}-i\epsilon\right)\,.
\end{equation}

The $i \epsilon$ and the $2 \xi$ factor in the argument of the logarithm are fixed by hand, extending the matching used at the one-loop level. These prescriptions are not needed for the resummation of $[\alpha_s \log^2 (\frac{\xi-x}{2\xi}-i\epsilon)]^n$ terms which we want to exhibit and are beyond the accuracy of our estimate.

\vspace{.2cm}

\paragraph*{The ladder diagram at order $\alpha_s^n$.} We now turn to the estimation of all $log^{2n} (x-\xi)$ terms. 
We again assume the strong ordering (\ref{kinematics_k}, \ref{kinematics_alpha}) in $k_{\perp}$ and $\alpha$. 
The distribution of the poles generates  nested integrals in $\beta_i$ 
\begin{equation}
\int_0^{\xi-x}d\beta_1\int_0^{\xi-x-\beta_1}d\beta_2\cdots\int_0^{\xi-x-\beta_1-\cdots -\beta_{n-1}}d\beta_n \,.
\end{equation}
The simplified numerator for the $n^{th}$ order box diagram  is  obtained by generalizing  the argument for the two-loop case. One gets
\begin{eqnarray}
&&\hspace{-.4cm}(Num)_n=-4s(x+\xi)^n\frac{2k_{\perp1}^2}{\beta_1}\bigg[1+\frac{2(x-\xi)}{\beta_1}\bigg] \frac{2k_{\perp 2}^2}{\beta_2}\\
&&\hspace{-.5cm}\bigg[1\!+\!\frac{2(\beta_1+x-\xi)}{\beta_2}\bigg] \!\!\cdots \!\frac{2k_{\perp n}^2}{\beta_n}\!\bigg[1\!+\!\frac{2(\beta_{n-1}+\!\cdots \!+\!\beta_1\!+\!x\!-\!\xi)}{\beta_n}\bigg] \nonumber\!,
\end{eqnarray}
and the denominators of  propagators are, for $i = 1 \cdots n$,
\begin{eqnarray}
L_i^2&=&\alpha_i(x+\xi)s \nonumber \\
R_i^2&=&-\underline{k}_{i}^2+\alpha_i(\beta_1+\cdots +\beta_i+x-\xi)s\,, \nonumber \\ 
S^2& =& -\underline{k}_{i}^2+(\beta_1+\cdots +\beta_i+x-\xi)s\,.
\end{eqnarray}
Using dimensional regularization and omitting scaleless integrals, the integral $I_n$  reads
\begin{eqnarray}
I_n=\hspace{-.1cm}\int_0^{\xi-x}\hspace{-.6cm}d\beta_1\hspace{-.1cm}\cdots\hspace{-.1cm}\int_0^{\xi-x-\cdots-\beta_{n-1}}\hspace{-.7cm}d\beta_n \hspace{-.1cm} \int_0^{\infty}\hspace{-.4cm}d^{d-2} \underline{k}_n\hspace{-.05cm}\cdots\hspace{-.15cm}\int_0^{\underline{k}_2^2}\hspace{-.3cm}d^{d-2}\underline{k}_1\,{\cal{I}}_n \,,\,\,
\end{eqnarray}
where
\begin{eqnarray}
&&\hspace{-.3cm}{\cal{I}}_n \!\!=\!\!(-1)^n\frac{4\, s(2\pi i)^n}{x-\xi}\frac{1}{\beta_1+x-\xi}\cdots\frac{1}{\beta_1\!+\!\cdots+\!\beta_{n-1}\!+\!x\!-\!\xi}\nonumber \\
&&\times \frac{1}{\underline{k}_1^2}\cdots\frac{1}{\underline{k}_n^2}\frac{1}{\underline{k}_n^2-(\beta_1+\cdots+\beta_n+x-\xi)s}\,.
\end{eqnarray}
The integrals over $\underline{k}_1 \cdots \underline{k}_n$  are performed again in a similar way as in the case of one- and two-loops, resulting in
\begin{equation}
\label{In-final}
I_n=-\frac{4}{x-\xi+i\epsilon}\frac{(2\pi i)^n}{(2n)!}\log^{2n}\left(\frac{\xi-x}{2\xi}-i\epsilon\right) \,,
\end{equation}
where the matching condition introduced in one-loop case is extended to $n-$loops.
\vspace{.2cm}

\paragraph*{The resummed formula.} 
Based on the results Eqs.~(\ref{I1-final}, \ref{I2-final}, \ref{In-final}), one can discuss the resummed formula for the complete amplitude.
By combining Eq.~(\ref{In-final}) with  Eq.~(\ref{def-Kn}), we get
\begin{eqnarray}
&&\sum_{n=0}^\infty K_n 
 =
\frac{e_q^2}{x-\xi + i \epsilon} \cosh
\left[ D \log \left( \frac{\xi -x}{2 \xi } - i \epsilon \right) \right]
 \\
&&= \frac{1}2 \frac{e_q^2}{x-\xi + i \epsilon}\left[ \left( \frac{\xi -x}{2 \xi } - i \epsilon \right)^D + \left( \frac{\xi -x}{2 \xi } - i \epsilon \right)^{-D} \right] \,.\nonumber
\label{Sumn-final}
\end{eqnarray}
Although the matching condition introduced at one-loop still does not fix the form of the resummed formula uniquely,  since it requires to go beyond double logarithmic accuracy, it is natural to propose the following two forms    
\begin{eqnarray}
&&\hspace{-.4 cm}(C_0+C_1)^{res}_1=\frac{e_q^2}{x-\xi+i\epsilon}\bigg\{
\cosh\bigg[D\log\bigg(\frac{\xi-x}{2\xi}-i\epsilon\bigg)\bigg]\nonumber \\
&&\hspace{-.4 cm}-\frac{D^2}{2}\bigg[9+3\frac{\xi-x}{x+\xi}\log\bigg(\frac{\xi-x}{2\xi}-i\epsilon \bigg)\bigg]\bigg\}-(x\rightarrow -x)\,,
\label{Res1}
\end{eqnarray}
with $D= \sqrt{\frac{\alpha_s C_F}{2\pi}}$ and 
\begin{eqnarray}
\label{Res2}
&&\hspace{-0.45 cm}(C_0+C_1)^{res}_2\!=\!\frac{e_q^2}{x-\xi+i\epsilon}
\cosh\bigg[D\log\bigg(\frac{\xi-x}{2\xi}\!-\!i\epsilon\bigg)\bigg] \\
&&\hspace{-0.4 cm}\times \bigg\{\!
1-\frac{D^2}{2}\bigg[9+3\frac{\xi-x}{x+\xi}\log\bigg(\frac{\xi-x}{2\xi}-i\epsilon\bigg)\!\bigg]\!\bigg\}
-(x\rightarrow -x) \,.\nonumber
\end{eqnarray}
These resummed formulas differ through logarithmic contributions which are beyond the precision of our study.

\vspace{.2cm}
\paragraph*{Phenomenological perspectives.}

\unitlength 1mm
\begin{figure}
\includegraphics[width=8.0cm]{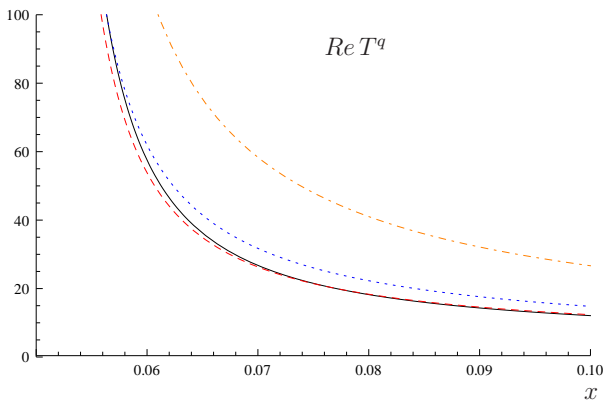}
\vspace{-2mm}
\caption{ $Re \, T^q$ as a function of $x$, for $\xi=0.05$ and $\mu_F=Q$:  coefficient function at LO (dot dashed-orange) and
 NLO  (solid-black), resummed formulas (\ref{Res1}) (dashed-red) and (\ref{Res2}) (dotted-blue). For simplicity, $e_q=1\,.$} 
\vspace{-4mm}
\label{Fig:Re}
\end{figure}

\begin{figure}
\vspace{.3cm}
\includegraphics[width=8.0cm]{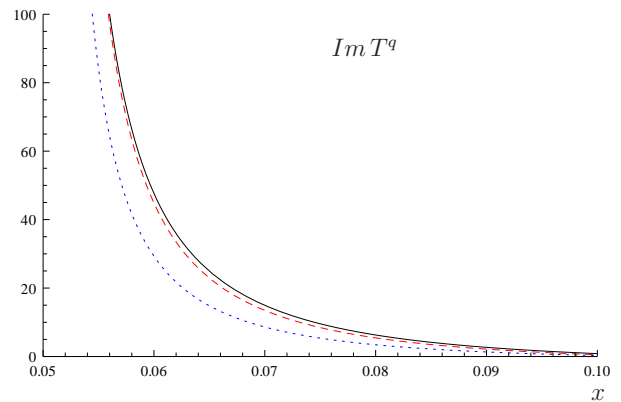}
\vspace{-2mm}
\caption{ $Im \, T^q$ as a function of $x$, for $\xi=0.05$ and $\mu_F=Q$:
NLO  coefficient function (solid-black), resummed formulas (\ref{Res1}) (dashed-red) and (\ref{Res2}) (dotted-blue), with $e_q=1\,.$} 
\vspace{-4mm}
\label{Fig:Im}
\end{figure}

In Figs.~\ref{Fig:Re} and \ref{Fig:Im}  we show the real and imaginary parts of the resummed coefficient function  ($\xi=0.05$ and $\alpha_s= 0.33$, corresponding to $Q=2$ GeV and $\Lambda_{QCD}= .2$ GeV)  together with the LO and NLO result, restricted to the DGLAP region. 
While the NLO corrections have sizable effects with respect to LO,
Fig.~\ref{Fig:Re}  shows that the higher order corrections when resummed in our way do not dramatically lead to any further change of the coefficient function. This hints to a stabilized perturbative expansion around $x=\pm\xi\,.$
Nevertheless, Figs.~\ref{Fig:Re} and \ref{Fig:Im} quantifies the sizable effects from the ambiguous choice of resummation formula (\ref{Res1}, \ref{Res2}), demanding a higher order analysis.

\vspace{.2cm}
\paragraph*{Conclusions.}
\noindent
We have demonstrated that resummation of soft-collinear gluon radiation effects can be performed in hard exclusive  reactions amplitudes. The resulting formula for coefficient function exhibits a different behaviour than both the Born order and the NLO results. 
A detailed investigation of the effect of resummation on observables such as  the magnitude of Compton form factors will be performed elsewhere.
A related but somewhat different expression should emerge in various hard exclusive reactions, such as exclusive meson electroproduction. The simplest extension of our formula will address the crossed case of timelike Compton scattering~\cite{MPSW}.

We did not study neither the effects of the running of $\alpha_s$ nor  the case of gluon GPD contributions which, although they begin at order $\alpha_s$ are expected to be important in the small $\xi$ regime to be accessible at high energies \cite{EIC-LHeC}.
It has been customary to perform resummation in terms of Mellin moments \cite{papers-resummation} . In our case, conformal Mellin moments may be more appropriate, and we will address this question in future work.
\vspace{.1cm}

We thank A.~H.~Mueller, G.~Sterman, O.~V.~Teryaev, J.~Wagner,    for useful discussions and correspondence.
This work is partly supported by  the Polish Grant NCN No DEC-2011/01/D/ST2/02069, the French-Polish collaboration agreement Polonium, the P2IO consortium
 and the Joint
Research Activity "Study of Strongly Interacting Matter"
(acronym HadronPhysics3, Grant Agreement n.283286) under the
Seventh Framework Programme of the European Community.


\end{document}